# Immersive Virtual Reality Environments for Embodied Learning of Engineering Students

Rafael Padilla Perez, Özgür Keleş Ph.D.
San Jose State University

## Abstract

Recent advancements in virtual reality (VR) technology have enabled the creation of immersive learning environments that provide engineering students with hands-on, interactive experiences. This paper presents a novel framework for virtual laboratory environments (VLEs) focused on embodied learning, specifically designed to teach concepts related to mechanical and materials engineering. Utilizing the principles of embodiment and congruency, these VR modules offer students the opportunity to engage physically with virtual specimens and machinery, thereby enhancing their understanding of complex topics through sensory immersion and kinesthetic interaction. Our framework employs an event-driven, directed-graph-based architecture developed with Unity 3D and C#, ensuring modularity and scalability. Students interact with the VR environment by performing tasks such as selecting and testing materials, which trigger various visual and haptic events to simulate real-world laboratory conditions. A pre-/post-test evaluation method was used to assess the educational effectiveness of these VR modules. Results demonstrated significant improvements in student comprehension and retention, with notable increases in test scores compared to traditional non-embodied VR methods. The implementation of these VLEs in a university setting highlighted their potential to democratize access to high-cost laboratory experiences, making engineering education more accessible and effective. By fostering a deeper connection between cognitive processes and physical actions, our VR framework not only enhances learning outcomes but also provides a template for future developments in VR-based education. Our study suggests that immersive VR environments can significantly improve the learning experience for engineering students, providing a powerful tool for educators to bridge the gap between theoretical knowledge and practical application.

## 1. Motivation and significance

Throughout recent history, rapid advances in computer graphics and rendering hardware have made it possible to create compact and affordable immersive technologies that interface with the human senses to create simulated realities through Virtual Reality. VR is typically defined as a complex media setup spanning a technological arrangement for sensory immersion, and a mechanism for the content representation of real and imagined worlds [18]. The main advantage of VR when compared to natural environments is it enables the observation of otherwise unobservable phenomena, reduces time and cost requirements for experimentation and testing for practical applications, and creates an opportunity for simulated environmental and social presence through virtual worlds at a higher degree of immersion [15]. The affordances of VR technology have transcended its typical use case associations with entertainment and video games. Statistically significant evidence supports an increase in research efforts and practical applications of these technologies within various fields. These include but are not limited to, nursing [1,4], K-12 education [2,3,4], STEM education [5,8,11-17,21,22], rehabilitation [6], and healthcare [20]. Additionally, it is worthwhile to recognize the growing trend of VR in college education for STEM-related fields and laboratory-style environments. In these settings, students can interact with equipment, samples, and procedures as they would in physical space, and even go beyond by studying more abstract concepts [1,4,8,11-15,21,22]. The design of VR environments is crucial to achieving learning goals effectively. Yet, the design of VR environments is not sufficiently documented, having limited free open-source examples.

Design frameworks for cognitively effective education have been explored, such as the use of the "Embodiment" principle of design described by Abrahamson and Lindgren [7], which describes the importance of creating activities that require perceptual senses and kinesthetic coordination, materials in orchestrated environments that allow somatic experiences, and facilitation to align the student's experience to that of experts in the field. The effects of this principle have been studied within Virtual Reality spaces [8], where three-dimensional virtual reality spaces were compared to their bi-dimensional counterparts such as video lectures on a flat screen. The problem with this testing method, is the high level of discrepancy in the educational context, as comparing a 2D to a 3D space conflicts with the scientific principle of comparative analysis, motive for other studies [9] to compare multiple levels of immersion within a virtual reality space and show a positive correlation between the level of immersion and spatial reasoning skills gained.

Another educational principle recently introduced is "Congruency," which as described by Latoschik and Wienrich [10] "is constituted by relations between the cues and the XR/VR experience itself." An example would be a student identically operating a complex machine in VR compared to its real-life counterpart. Following this rationale aiming to maximize Embodiment and Congruency, while considering the principle of comparative analysis we've proposed a novel Virtual Laboratory Environment (VLE) framework. The framework presented here is applied within the domain of materials and mechanical engineering to supplement student education in tensile testing practices, and the physical behavior of Poisson's ratio coefficients in different materials. These types of VR modules can be extended to cover whole-semester courses or to be used as supplemental modules for enhance learning, engagement, and excitement.

## 2. Software description and functionality

The goal of this framework is not only to provide a simulation environment for education in VR but also to evaluate its educational effectiveness. Traditional VR comparative analysis methods, such as our previous research [22] have a high discrepancy in testing environments, as the effectiveness of learning through 2D slideshows and lectures is compared with their modeled versions in virtual reality, yielding results that contradict the scientific principle of comparative analysis. To address this, our method was tested using VR exclusively and quantitatively through a pre-/post-test evaluation. In our case, a set of two laboratories where the majority of interactions are embodied and involve kinesthetic motion was compared to an earlier version [22], where they were done through laser selection and the lack of walking capability (teleportation).

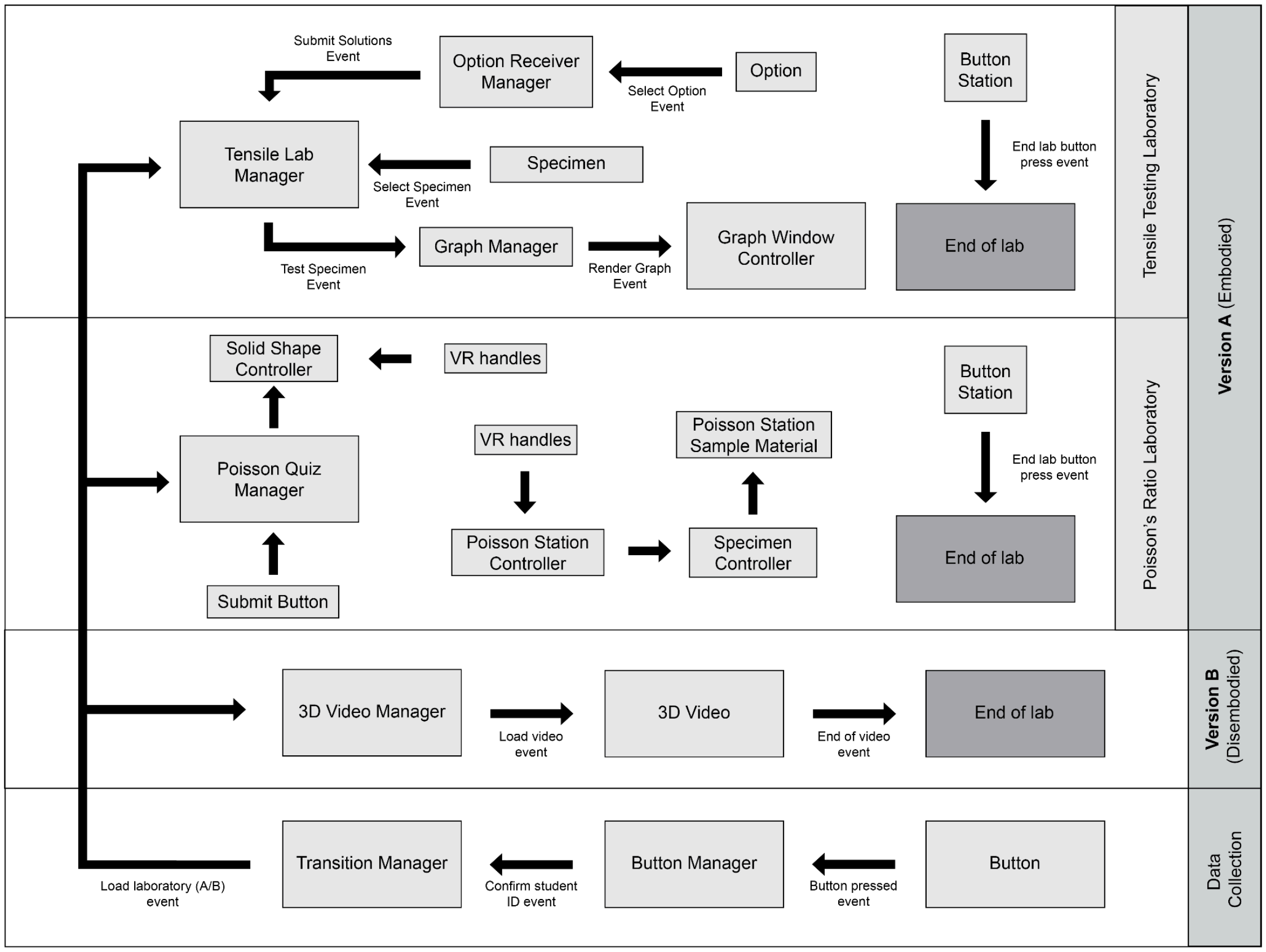

Figure 1. Directed graph software architecture showing version A of the VLE with all events and observers for Data Collection, Tensile Testing, and Poisson's Ratio modules (Note version B was not tested in this study).

Fig. 1 describes an event-driven directed-graph-based architecture, where specific components within the virtual reality space trigger different visual and haptic events upon interaction. These applications, developed with C# and Unity 3D leveraging the XR interaction toolkit, and the OpenXR framework for the Meta Quest 2 were designed with modularity as a foundational design pillar. The system's modular nature arises from the replaceability potential of any given feature, as any VR interaction such as a button press, lever pull, or a hand gesture can be utilized to trigger an event of any kind.

## 2.1 Software architecture

In software engineering, there are common programming paradigms and design patterns that allow for abstraction and scalability, one of which is the Observer Pattern. This pattern creates a mechanism for methods to execute upon specific events, therefore we can categorize behaviors into "events," which are triggers from world state changes, and "observers," which execute as a result of events triggering. This pattern is at the core of our VLE, as it allows for interchangeable interaction types using the same laboratory. An example would be the "Select Specimen Event," which occurs between the Specimen (the selected event) and Tensile Lab Manager (the observer) as seen in Fig. 1. Whenever the student selects one of the three specimens in the laboratory (a polymer, a metal, and a ceramic) by grabbing it with their hand, the event triggers an action in subscribed methods within the Tensile Lab Manager, such as saving the state of that specimen as selected ( see Fig. 1 "Select Specimen Event") for further logic processing. Events are highly extensible, as they can trigger behaviors such as showing a video, giving the user haptic feedback with a rumble, or playing a sound.

Our software presents modules that exhibit the application of the congruent and embodied principles previously discussed. Fig. 2 demonstrates the first module, which focuses on tensile testing. a) Student begins by getting familiar with the VR environment (hand interactions, locomotion) by walking a short path and entering their student ID in a pin pad using the "Button Pressed Event" described in Fig. 1. b) Student is instructed using the self-guiding system to grab a specimen. c) Student grabs a specimen, triggering the "Select Specimen Event", giving rumble feedback and parenting the specimen's world position to the student's hand. d) Student aligns the sample to tensile tester grips. e) Stress/strain graph with physical data is generated. f) Student identifies Yield, Tensile, and Fracture strength. g) Student removes a tested sample from the tensile testing machine. h) Student identifies the material type based on its stress/strain graph.

In Fig. 3, the next module which highlights the properties of Poisson’s Ratio is illustrated. a) The student is shown three "stations" with samples exhibiting different Poisson ratios. b) Student applies a positive pressure as a function of the separation between the suspended handles shown, a VR interaction we call "handle diversion" which in this case, updates the strain-strain curve and deforms the material. c) Student applies negative pressure, deforming the sample.

The student, after interacting with all three stations shown in a) triggers the next portion of the laboratory shown in d), where they solve a learning assessment. e) The student selects a response for the resulting deformation of a material under compressive stress with a Poisson ratio of 0 in its X-axis. f) The student controls his solution using the handle diversion interaction. Once the student is satisfied with all their results, they press the green "Check" button shown in d), which highlights correct solutions in green and incorrect ones in red, a quizzing mechanic also present in the tensile laboratory shown in Fig. 2.

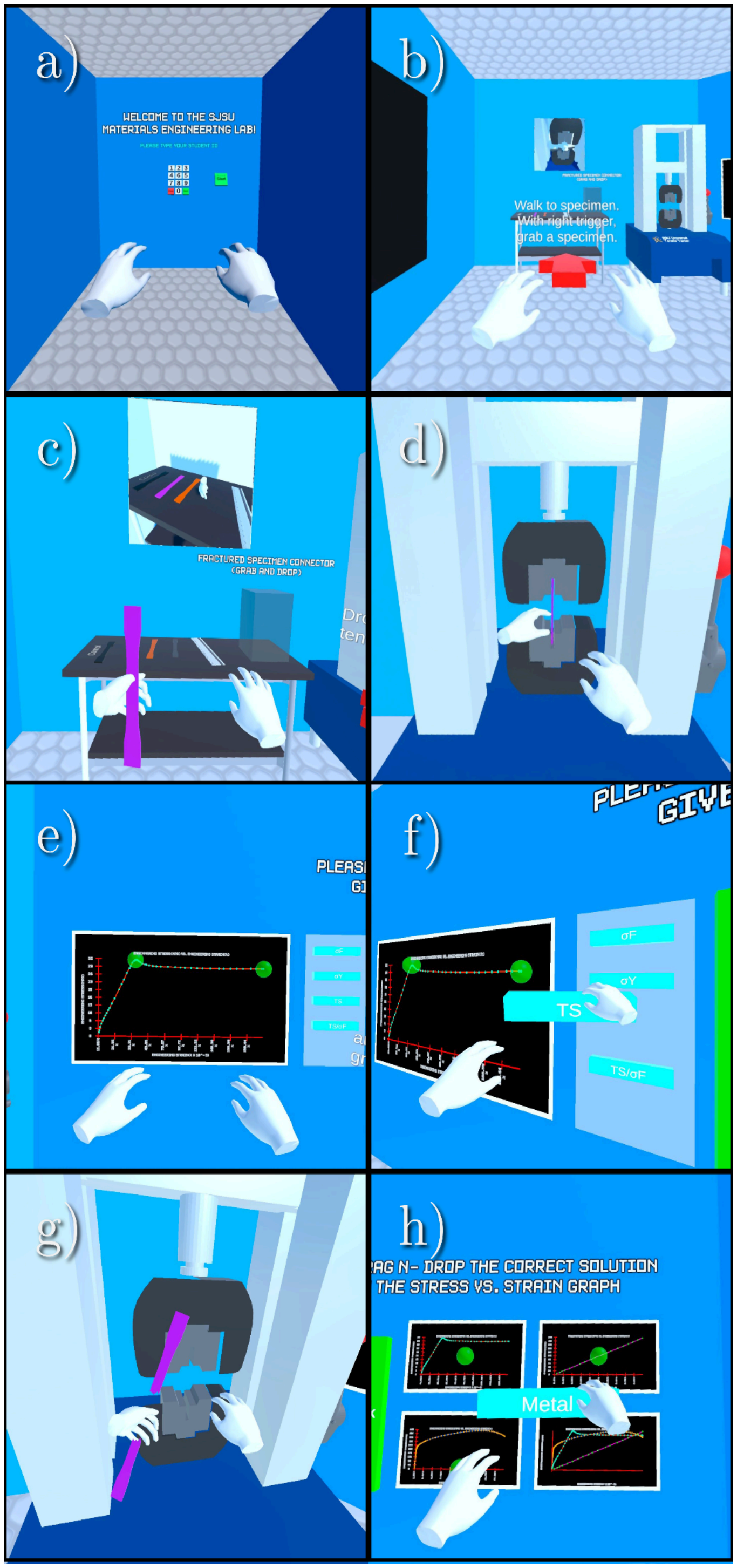

Figure 2. Tensile testing VLE module a) scene where students enter their IDs and start the VLE experience, b) overview of the tensile testing VLE, c) standard-size tensile testing specimen, d) student using hands to put the sample between the grips, e) tensile test result with green areas

requesting to put the related mechanical property by taking the boxes shown in f), g) student takes out the specimen, and h) final quiz is shown where the student selects material type and puts in the correct stress-strain diagram.

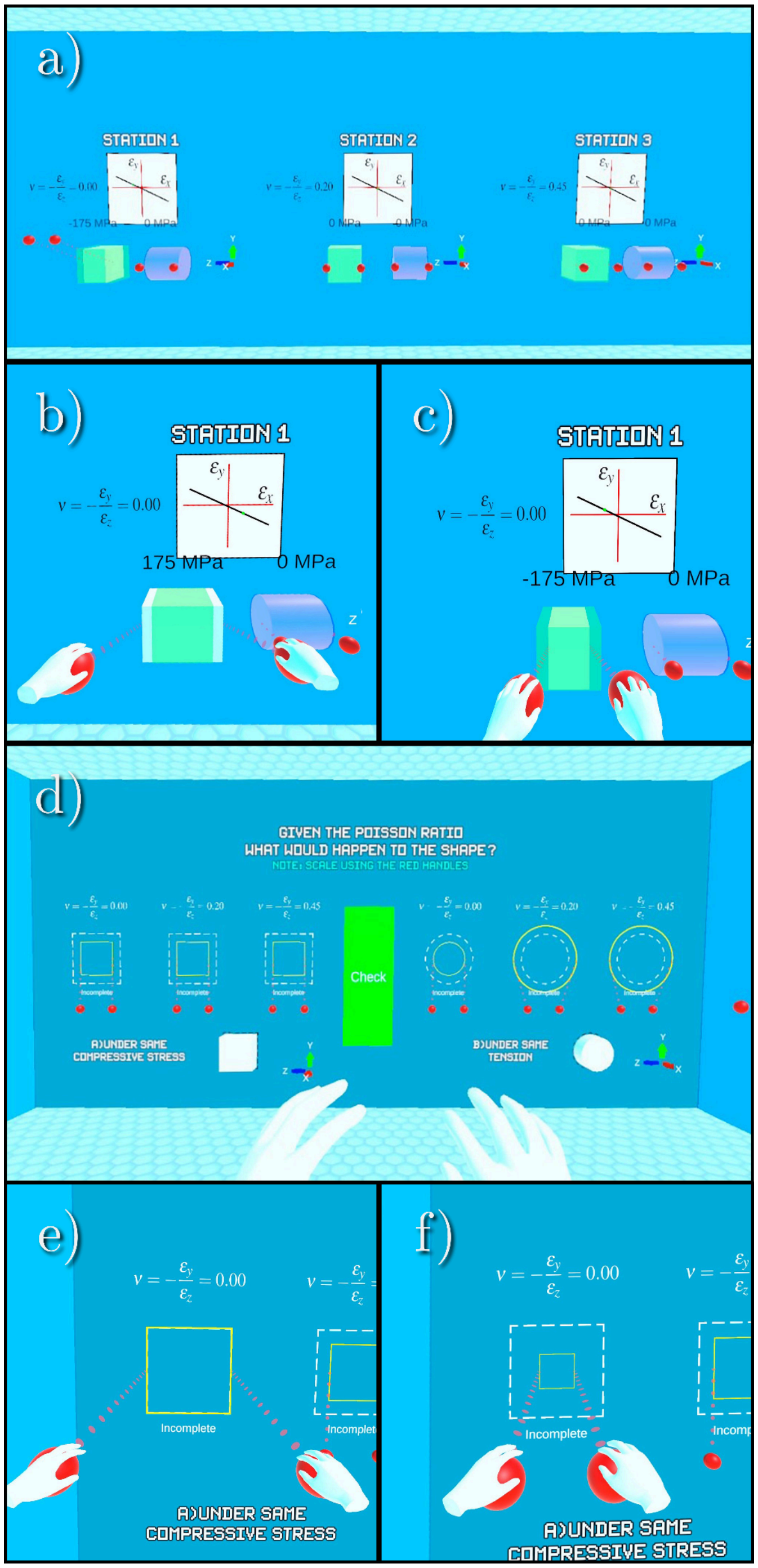

Figure 3. Poisson's ratio VLE module: a) general overview of the VLE showing three materials with different Poisson’s ratio, the rectangular and cylindrical specimens are the same material that shows the effect of lateral contraction and expansion under tension or compression. The tension and compression are applied by students who handle the red spheres and b) expand them to apply tension and c) contract them to apply compression. d) a unique in-VLE quiz asking students to find correct shape change for a cube and cylinder under tension or compression (e and f). Students need to find the correct answers to proceed to the next module.

**3. Illustrative examples**

To illustrate the software's capabilities, a supplementary video [https://youtu.be/NBAVW_nH-1w] provides a full demonstration of the Tensile Testing and Poisson's Ratio modules from the student's perspective. It showcases the interactions between the student and the VLE modules, presenting a complete walkthrough of the steps required to receive credit for the laboratory modules. The tensile testing module is modeled after a popular materials engineering laboratory assessment at a University level, where students must characterize different material samples by analyzing their stress/strain graphs using a tensile testing machine. The Poisson's ratio VLE however, presents an interactive demonstration of how cubic and cylindrical samples deform after stress or strain is applied to them and how Poisson's ratio affects deformation along distinct axes. This phenomenon is challenging to observe physically, let alone interact with using touch, which is why it was implemented with VR. In both modules, students follow a self-guided, self-paced system supported by text prompts and a directional arrow guiding them task-by-task, until all requisites of the module are completed successfully.

**4. Impact**

Direct comparisons between fully embodied, limited embodied, and non-embodied virtual learning environments can be tested on different student groups. This can be done as a quick one-stop VR experience for materials, mechanical, aerospace, civil, and biomedical engineers as mechanical testing and material properties cover a wide range of disciplines. This endeavor was the result of the complementary effort between Software Engineering undergraduate students and Faculty from the Chemical and Materials Engineering Department at San Jose State University. Fig. 4 illustrates students interacting with the VLEs during a lecture session of the MATE25 (Introduction to Materials Engineering) course on campus using our VR teaching framework. The cooperation of the different disciplines involved in this project resulted in not only a quantitatively improved version of previous software but also the creation of a new testing methodology which we believe will strongly influence the educational quality of forthcoming VR modules for use in learning environments at schools and universities.

This software may be utilized directly by instructors aiming to effectively teach the concepts pertaining to Tensile Testing and Poisson's Ratio using VR or may be uniquely adapted to specific subjects within STEM and beyond. The two VR modules presented will be used moving forward at San Jose State University as an instructional tool for the MATE25 course, impacting four, fifty-student sections (totaling 200+ students) per semester. Furthermore, this software will facilitate research on which VR interaction and embodiment methods are most effective for education as further VR research and modules continue in development.

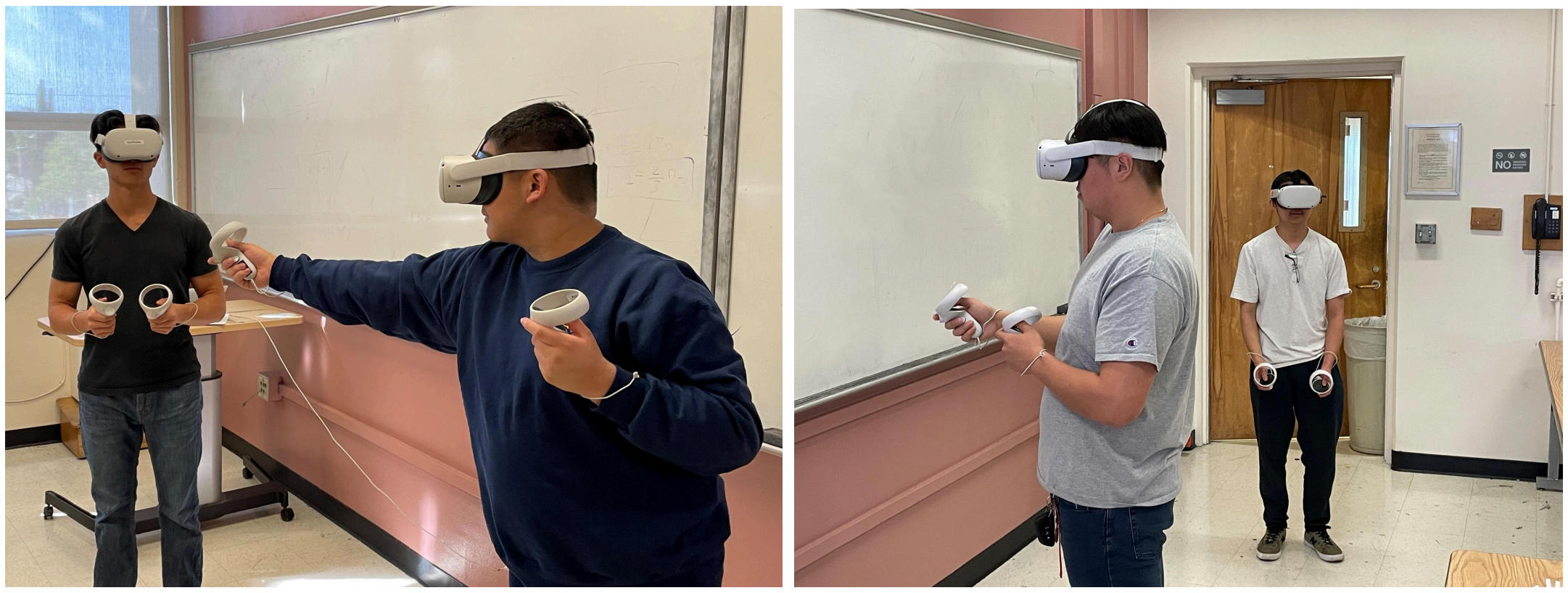

Figure 4. Students using the immersive  VLEs during a class.

**5. Student trial**

Students who had basic knowledge of physics from materials, mechanical, chemical, and aerospace engineering experienced the VLEs. A pre-test was given to the students, followed by the self-paced VLE experience, and a post-test (see Appendix A for the pre and post-test questions). Students had ~5 min. to complete both tests.

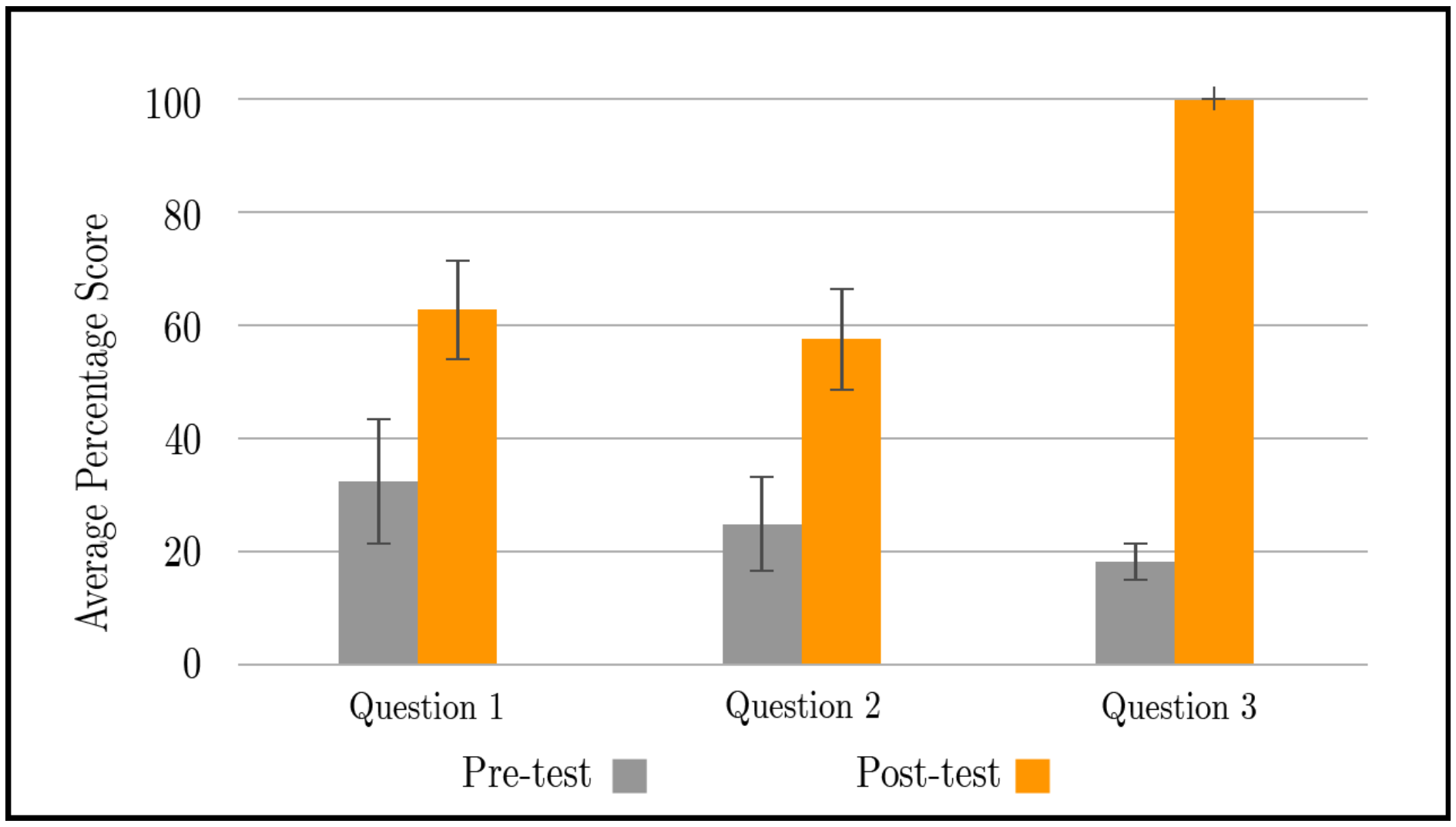

Figure 5. Average scores of pre- and post-tests for questions 1-3 (error bars represent standard deviation).

Students had higher scores for all the questions after the VLE experience. There was a 49%, 133%, and 65% increase in scores for questions 1, 2, and 3, respectively. In a previous study on the same tensile testing and Poisson's ratio subjects, students had a version of the VLEs designed without embodiment and congruency principles. The same testing approach resulted in 3% decrease, 78% increase, and 47% increase for questions 1, 2, and 3, respectively [22]. Note that, embodied VLE sessions took about 10 minutes per student; whereas, the previous VLE experience was 20 to 30 minutes.

## 6. Conclusion

Virtual reality provides simulation capacity for educational environments that are expensive to replicate, such as those within the realm of engineering that require heavy machinery, specialized tools, and high-cost materials. It democratizes learning opportunities by making them more accessible to more academic institutions, at a fraction of the cost with the added benefits of enhanced stundent learning, student engagement and excitement. The scalability of the VR software also enables significant impact on large number of learning in the world, while collection one-of-a-kind learning data. This creation of big data will enhance future artificially intelligent teaching approaches. Finally, in the current design, a novel split-testing format was also developed, which introduced two versions of the laboratories, one with a high level of embodiment and congruency ("Version A, Fig. 1") where students interact with simulated objects and perform a learning assessment, and another where they observe a demonstration of the laboratory in a 3D video within the same virtual reality environment ("Version B, Fig. 1"), with a lower level of immersion. Although we did not test the 3D video environment, this structure enables direct comparison between VR interaction, and non-interaction within the same VR environment.

**Funding**
Partial funding for this work was provided by the College of Engineering, San Jose State University. Partial funding for this work was provided by the U.S. National Science Foundation CAREER Award No. 2145604.

**Appendix A**
Questions used in pre and post-tests.

# Pre-test question 1

1) You are given three same sized cubes with Poisson's ratio of 0.48, 0.22, and 0. Draw three more squares representing the bottom surface of these cubes **under tension**. The squares given below represent the bottom of the cubes under no stress, clearly show the relative sizes of your squares.

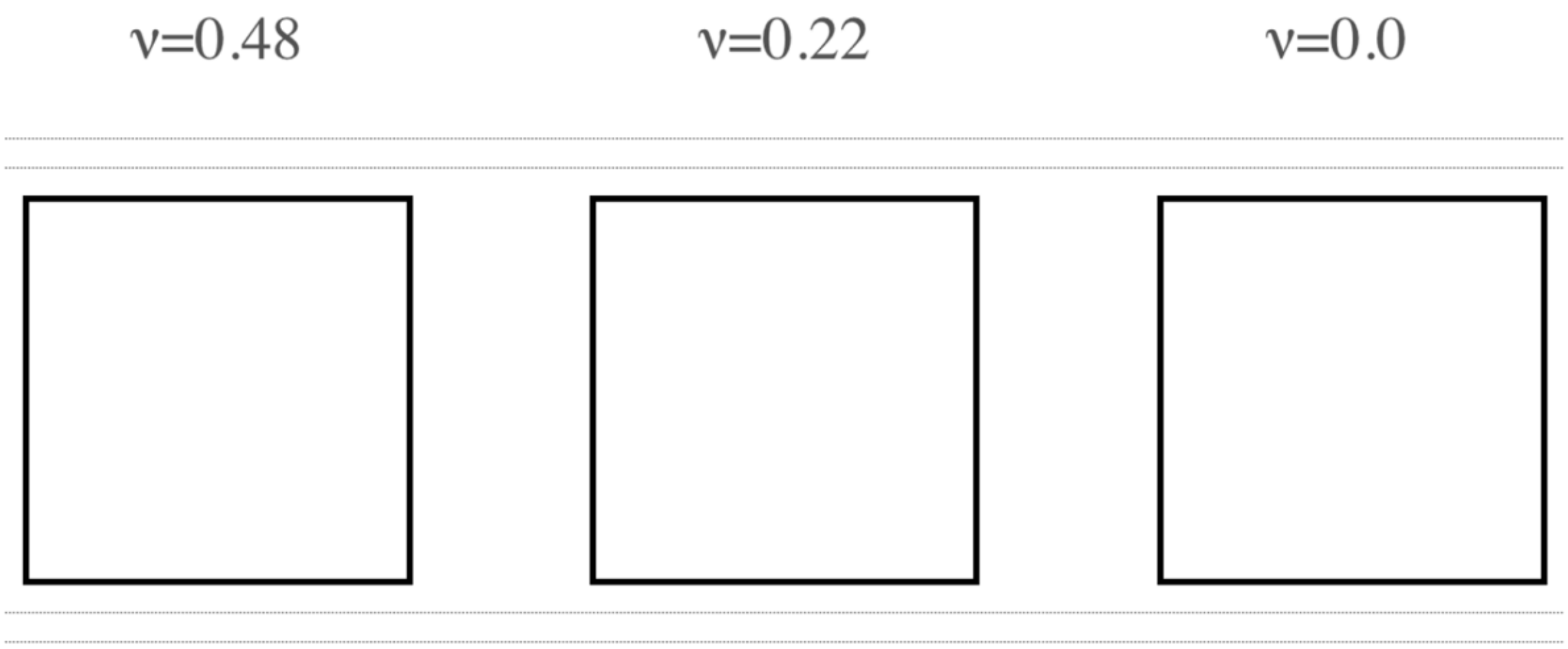

# Post-test question 1

1) You are given three same sized cylinders with Poisson's ratio of 0.48, 0.22, and 0. Draw three more circles representing the bottom surface of these cylinders **under compression**. The circles given below represent the bottom of the cylinders under no stress, clearly show the relative sizes of your circles.

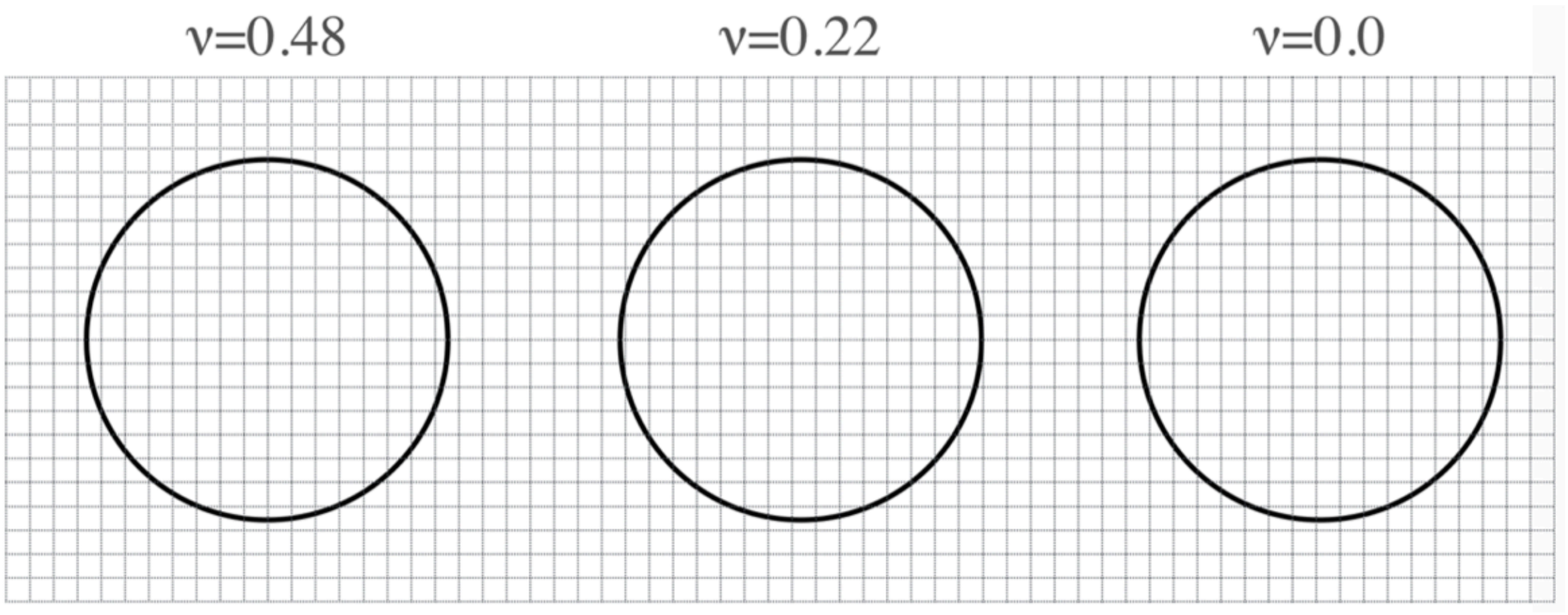

# Pre- and post-test questions 2 and 3

2) You are given stress-strain diagrams of three materials: (a), (b), and (c). Clearly show the tensile strength (TS) and yield strength ($\sigma_y$) of these materials on the graph. Identify the general material type of these materials.

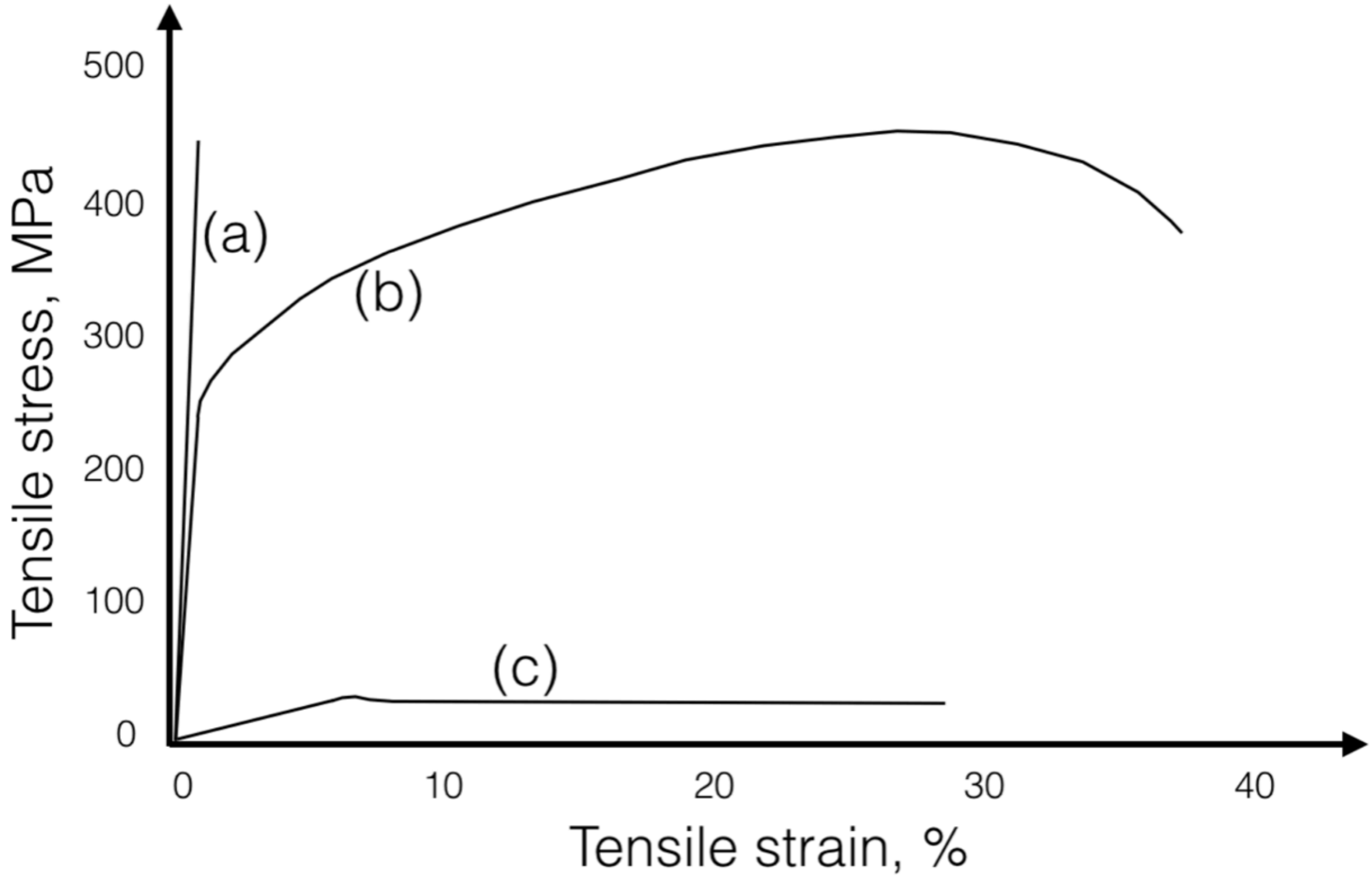

3)
Material type for:

a)

b)

c)